# A new approach for a proof that P is NP

Malay Dutta

Ex-Professor

Indian Institute of Information Technology Guwahati

and Anjana K. Mahanta

Department of Computer Science Gauhati University

## ABSTRACT

In this paper we propose a new approach for developing a proof that P=NP. We propose to use a polynomial-time reduction of a NP-complete problem to Linear Programming. Earlier such attempts used polynomial-time transformation which is a special form of reduction that uses a subroutine for the easier Linear Programming Problem only once. We use multiple calls to the subroutine increasing considerably the effectiveness of the reduction. Further the NP-complete problem we choose is also unusual. We define a special kind of acyclic directed graph which we call a time graph. We define Hamiltonian time paths in such graphs and also the Hamiltonian Time Path problem (HTPATH) and prove that it is NP-complete. We then state a conjecture whose proof will immediately lead to a polynomial-time algorithm for this problem proving P=NP.

**1. Introduction :** The P=NP problem is one of the most important open problems in Theoretical Computer Science today and it also has many significant connections in the world of applications. The problem concerns itself with decision problems i.e. computational problems which have only yes-no answers. P is the class of decision problems which have polynomial-time algorithms and NP is the class of decision problems whose inputs suspected to have yes answers can be checked in polynomial-time. It is easy to show that P is contained in NP. But it is not known whether every problem in NP is in P. A class of problems known as NP-complete problems have been found which are the hardest problems in NP in the sense that if any NP-complete problem is found to be in P then NP will become P. Many problems in the application world have been proved to be NP-complete, and since they are important, various inexact methods have been developed to handle them.

One important technique of developing an algorithm for a difficult problem is a polynomial-time reduction to an easier problem. The Linear Programming Problem was not known to have a polynomial-time algorithm for a long time but was very useful in designing good algorithms for many problems in applications. So when a polynomial-time algorithm for Linear Programming was found, a lot of efforts were made to reduce in polynomial-time some NP-complete problem to Linear Programming (more precisely to Linear Programming Feasibility which is a decision problem). But most of these attempts used a form of reduction, called a polynomial-time transformation (Definition 15.3 of [2]) which uses just one call to a subroutine containing the algorithm for the easier problem. Such a subroutine for a problem in P is just like a small hammer with which we want to demolish the structure of a big building (the NP-complete problem). It is intuitively obvious that we should call such a routine many times (of course a number of times bounded by a polynomial). Each time we will achieve a relatively small objective. This is the approach we have taken in this paper.

We have also taken a special choice of the NP-complete problem to be reduced to an easier problem. The Hamiltonian cycle problem on digraphs is a NP-complete problem but in an attempt for a reduction to Linear Programming, the possibilities of having disjoint cycles give problems. We start with the Hamiltonian Cycle problem on digraphs and we introduce a time dimension in the vertices and consider the vertices as (i,t) where i represents the original vertex number and t represents a time layer. We also split one of the vertices (say the vertex 0) into two – (0,0) as a source vertex and (0,n+1) as a destination vertex. The edges are now from one layer to the next layer and the graphs become acyclic. The Hamiltonian cycles in the original graph get mapped to what we will call Hamiltonian time paths and the graph will be called a time graph of order n if the original digraph had n+1 vertices counting the vertex 0 which was split into two.

Hence we define a complete time graph $K_n^T$ of order n as a layered directed graph with a single source vertex (0,0), a single destination vertex (0,n+1) and $n^2$ interior vertices (i,t) (1≤i≤n;1≤t≤n). Edges are $e_{0j0}$ from (0,0) to (j,1) for 1≤j≤n, $e_{ijt}$ from (i,t) to (j,t+1) for (1≤i,j≤n (i≠j), 1≤t≤n-1) and finally $e_{i0n}$ from (i,n) to (0,n+1) for 1≤i≤n. This definition of $K_n^T$ is based on the notion of a time dependent traveling salesman problem [1]. We will be concerned here with the possible paths that a tourist may take within n + 1 cities 0, 1, 2, … , n. Suppose the tourist starts at city 0 on day 0, travels through city $p_i$ on day i for 1 ≤ i ≤ n ($p_i ≠ p_{i+1}$), 1≤$p_i$≤n and finally returns to city 0 on day n + 1. We can formulate the path that the tourist takes to be a path in the special directed graph $K_n^T$. Keeping this in mind, for a vertex (i,t) i is called the city number and t is called the day number. For the edge $e_{ijt}$, (i,t) is the source, (j,t+1) is the destination and t is the time layer consisting of the interval between day t and day t+1. For the edge $e_{ijt}$, i must be 0 if t=0 and j must be 0 if t=n. A time path is a path from (0,0) to (0,n+1). It must go through the vertices (p(t), t) for 1≤t≤n, where p() is a function from {1,2,…,n} to {1,2,…,n}. If p() is a permutation then the path is called a Hamiltonian time path (htp). The characteristic function of a htp will be called a Hamiltonian time flow (denoted by a 'htf') and is a vector in the space of rational-valued functions on the set $E(K_n^T)$ of edges of $K_n^T$, which forms a vector space under point-wise operations. For a function f on $E(K_n^T)$, support(f) will be the set of edges where f is non-zero. Support(f) is empty iff f = 0.

A time graph G of order n is a subgraph of $K_n^T$ with the same vertices as that of $K_n^T$ and the set of edges E(G) which is a subset of $E(K_n^T)$. A htf in G will mean a htf with support a subset of E(G). The vector space of functions generated by the htf's in G will be denoted by H(G). For f in H(G), the sum of the values of f over the edges in any time layer will be the same for every layer and this quantity will be called the flow of f. A time graph G is called Hamiltonian if there is a htf in G. The Hamiltonian time path problem (HAMTPATH) is the problem : Given a time graph G of order n, is G Hamiltonian ?

In section 2, we will prove that HAMTPATH is NP-complete. In section 3 given a time graph G and an edge e in G, we define a Linear Programming problem LP(G,e) which is infeasible if there cannot be a htp in G passing through e. Based on this definition we classify some edges of G as useless. In section 4 we state our conjecture. In the concluding section 5 we prove that if this conjecture is true we get a polynomial-time algorithm for HAMTPATH, which will lead to the proof that P=NP. In this algorithm, HAMTPATH gets reduced to Linear Programming Feasibility with multiple (but a polynomial number of) calls to a subroutine determining the feasibility of a Linear Programming Problem. This gives a polynomial-time reduction of HAMTPATH to Linear Programming Feasibility problem in the sense of Definition 15.2 of [2] (not a polynomial-time transformation as per Definition 15.3 of [2]).

**2.    NP-completeness of the HAMTPATH problem :** In [3], the HAMPATH problem is defined in Section 7.3 as the problem of deciding whether a directed graph contains a Hamiltonian path connecting two specified source and terminal nodes s and t. It is shown in Theorem 7.46 of [3] that HAMPATH is NP-complete. We shall prove here that HAMPATH can be reduced to HAMTPATH in $O(n^3)$ time proving that HAMTPATH is NP-complete.

Theorem 2.1 : HAMPATH can be reduced to HAMTPATH in $O(n^3)$ time.
Proof :  Given a directed graph D with the set of vertices {s,t,1,2,…,n} with s as a designated source vertex, t as a designated terminal vertex and the set of edges E, we construct a time graph G of order n as follows :

For every t=1,2,…n-1; $e_{ijt}$ is an edge in G iff (i,j) is in E. For every j=1,2,…,n; $e_{0j0}$ is an edge in G iff (s,j) is in E. For every i=1,2,…,n; $e_{i0n}$ is an edge in G iff (i,t) is in E. It is then easy to prove that D has a Hamiltonian path from s to t iff G has a Hamiltonian time path. Hence the construction of G given D is a reduction from HAMPATH to HAMTPATH. Also given D, the construction of G can be carried out in $O(n^3)$ time. This proves the theorem and we can conclude that HAMTPATH is NP-complete.

3. **Some Preliminary Definitions and Results :** Let G be a time graph of order n and e be an edge in G. We define the LP problem LP(G,e) : For a function f on E(G) satisfying the conditions :

$\sum_j f(e_{0j0}) = 1$                      (1)

$\sum_j f(e_{ijt}) - \sum_{j'} f(e_{j'it-1}) = 0$    for 1≤i,t≤n        (2)

$\sum_{jt} f(e_{ijt}) = 1$    for 1≤i≤n        (3)

$\sum_i f(e_{i0n}) = 1$    for 1≤i≤n        (4)

$f(e) = 1$                             (5)

$f(e') \geq 0$   for every e' in E(G          (6)

minimize <0,f>)

Any htf f in G with f(e) = 1 is a feasible solution of LP(G,e). Hence if LP(G,e) is infeasible there cannot be a htp in G passing through e. Such an edge will be called useless in G and the removal of an useless edge from G does not affect the set of htp's in G.

4. **Our Conjecture :** We now state our conjecture :

Conjecture 4.1 : A time graph with a nonempty set of edges where no edge is useless, is Hamiltonian.

Suppose this conjecture is not true and G is a non-Hamiltonian time graph with a nonempty set of edges where no edge is useless. We start with some edge e in G. Since e is not useless, there is a feasible solution f to LP(G,e). The solution cannot represent a htf since G is not Hamiltonian. Hence there are edges e1, e2 in G such that f(e1) and f(e2) < 1. We shall call e1 and e2 to be children of e. But in turn e1 and e2 are not useless. Hence each of them will have a pair of children. If we can somehow show that this cannot go on indefinitely (because of the finite number of edges), then the conjecture will get proved. We have tried our best to get a counter-example to the conjecture but could not get one. This has given us a faint hope that the conjecture may be true. If this conjecture is true, we shall show in the next section that P = NP will get proved. By the way the conjecture opens a new problem which the reader can try as a challenge.

5. **Consequence of the truth of this Conjecture :** Suppose Conjecture 4.1 is true. Then the obvious way of checking whether a given time graph G is Hamiltonian will be to remove all the useless edges of G. If all the edges get removed then G is not Hamiltonian. If we are left with a nonempty set of edges with no useless edge then the resulting graph G' is Hamiltonian by Conjecture 4.1. Since the edge-set of G' is contained in the edge-set of G, G is Hamiltonian.

However we have to be careful in the order of the removal of the edges. For example for the time graph of order 5 with edges $e_{010}$, $e_{121}$, $e_{131}$, $e_{242}$, $e_{352}$, $e_{453}$, $e_{543}$, $e_{524}$, $e_{434}$, $e_{205}$, $e_{305}$, if we try to remove $e_{010}$ first, it will not be removed. After that if we try the other edges then all of them will get removed since these are useless edges. As a result we will remain with the single edge $e_{010}$ and apparently it will violate the conjecture and declare the graph to be Hamiltonian. The correct way to check edges for removal will be to repeat the process from the beginning as soon as an edge is found to be useless. Thus using Conjecture 4.1 we get the following algorithm for the HAMTPATH problem.

Algorithm 5.1 :

Given a time graph G :
1) G' = G
2) If G' has no edges return "Not Hamiltonian"
3) For every e in G'
4) If LP(G',e) is infeasible remove e from G' and go to 2)
5) End for 3
6) If G' has some edge left return "Hamiltonian"

There are O($n^3$) edges in G. For every edge in G, LP(G',e) is invoked at most once. The size of LP(G',e) is bounded by a polynomial in n and the complexity of finding the feasibility of LP(G',e) is bounded by a polynomial in the size of LP(G',e). Hence Algorithm 5.1 runs in time bounded by a polynomial in n. Since Algorithm 5.1 solves the HAMTPATH problem which is NP-complete, we can conclude that P = NP provided Conjecture 5.1 is true.